\pdfoutput=1 
\documentclass{JINST}
\usepackage{amsmath}
\usepackage{textcomp}

\title{Potential of Thin Films for use in Charged Particle Tracking Detectors}

\author{
J. Metcalfe$^a$\thanks{Corresponding author.}~,
I. Mejia$^b$,
J. Murphy$^b$, 
M. Quevedo$^b$,
L. Smith$^b$,
J. Alvarado$^c$,
B. Gnade$^b$, 
and H. Takai$^a$\\

\\
 \llap{$^a$}Brookhaven National Laboratory,\\
  PO Box 5000, Upton, NY 11973, USA\\
\llap{$^b$}University of Texas, Dallas,\\
  800 W Campbell Rd, Richardson, TX 75080, USA\\
\llap{$^c$}Benem$\acute{e}$rita Universidad Aut$\acute{o}$noma de Puebla,\\
  Calle 4 Sur 104, Centro Historico\\ 72000 Heroica Puebla de Zaragoza, PUE, Mexico\\  
E-mail: \email{jessica.metcalfe@gmail.com}}

\abstract{Thin Film technology has widespread applications in everyday electronics, notably Liquid Crystal Display screens, solar cells, and organic light emitting diodes. We explore the potential of this technology as charged particle radiation tracking detectors for use in High Energy Physics experiments such as those at the Large Hadron Collider or the Relativistic Heavy Ion Collider. Through modern fabrication techniques, a host of semiconductor materials are available to construct thin, flexible detectors with integrated electronics with pixel sizes on the order of a few microns. We review the material properties of promising candidates, discuss the potential benefits and challenges associated with this technology, and review previously demonstrated applicability as a neutron detector.}


\begin{document}

\section{Introduction}
Thin Film (TF) technology is rapidly advancing due to widespread commercial application in consumer electronics such as liquid crystal displays (LCD) and organic light emitting diodes used in TVs, computer monitors, tablets, and cell phones. TF's boast benefits such as optical transparency, mechanical flexibility, high spatial resolution, large area coverage, and low cost over traditional silicon based semiconductor technology. TF technology uses crystalline growth techniques to layer materials. Layers of particle detection material with amplification electronics can be used to create a particle radiation detector. The advantages of a radiation detector made with this type of technology include single piece large area device (on the order of a few m$^2$), high resolution ($<$10 $\mu$m), low cost ($\times$100 less than Si-CMOS), low mass, and high curvature for a cylindrical, edgeless design \cite{street,gnade,mejia}. 

\subsection{TF fabrication}
The fabrication process uses chemical bath deposition and close-space sublimation techniques on a substrate material to produce thin films with a high degree of precision. Here, the crystalline structure is grown in layers rather the using drilling and etching techniques standard in traditional silicon fabrication. The TF processing is much less expensive. Recently, films as thick as 200 $\mu$m--as thick as current silicon bulk--was achieved. \cite{thickFilms}

\section{TF Design Concept}

\subsection{Design Goals}  
TF technology opens the door to new materials to replace traditional silicon detectors. The fabrication can be done on flexible substrates such as organic polymers and plastics for flexible detectors. It is conceivable large sheets could be rolled around the beam pipe eliminating the dead areas currently around each silicon sensor. The sensors are very thin, between 5 and 50 microns. Including the integrated electronics, the expected radiation lengths of a single TF detector layer will be only a few percent compared to that of a silicon detector due to the smaller radiation length of other potential semi-conductor materials and its thinness. The cost of a TF detector is also orders of magnitude less than traditional silicon. For example, the cost estimate for polycrystalline CdTe charged particle detector is estimated at $<$\$10 per m$^2$ for a 2.5 $\mu$m thick film. There is also a large saving in the power to bias a TF detector--around 2 V (for a 6 $\mu$m thick film) is required for the bias voltage compared to approximately 100 V for silicon sensors and approximately 5-10 V is anticipated for the ``front-end" electronics. Many of the materials suggested also work well at room temperature reducing the need for cooling infrastructure. The myriad of potential materials in TF processing also allows the technology to be optimized for different operational needs making it more adaptable.  

The goal of a TF based tracking detector is to meet the current performance goals of tracking detectors such as ATLAS or CMS at the LHC at CERN. In order to achieve this some target performance points are established. These include charge yield on the order of 1,000 to 10,000 electrons, an energy resolution of 5-10\%, position resolution better than 10 $\mu$m, and timing resolution on the order of 10 to 100 ns. The most challenging points expected are the charge amplitude of a signal minimum ionizing particle (MIP) and the timing resolution. These will be addressed in Sections~\ref{sec:charge} and \ref{sec:timing}.

\subsubsection{Charge Generation}\label{sec:charge}
The key elements of signal collection in a charged particle detector can be broken down into three main components--the amount of charge deposited into the material, the subsequent charge that is generated, and the percent of charge that is read-out. The amount of charge deposited is proportional to the atomic number, Z; atomic weight, A; and the density of the material, $\rho$, as shown in Equation~\ref{eqn:dEdx}:
\begin{equation}
-\frac{dE}{dx} = \frac{Z\rho}{A}\frac{Kz^2}{\beta^2}\ln[\frac{2m_ec^2\beta^2}{I^2(1-\beta^2}-\beta^2]
~\propto ~ \frac{Z\rho}{A}
\label{eqn:dEdx}
\end{equation}
where $c$ is the speed of light, $m_e$ is the electron mass, $z$ is the charge of the incident particle, $\beta$ is velocity of the incident particle divided by $c$, and $I$ is the mean excitation potential, which can be approximated by 10 eV$\cdot Z$. The value $K$ is a constant given by:
\begin{equation}
K = 2\pi N_A r_e^2 m_e c^2 = 0.1535~ [\frac{MeV \cdot cm^2}{g}]
\end{equation}
where $N_A$ is the Avagadro Number constant and $r_e$ is the classical radius of an electron:
\begin{equation}
r_e = \frac{1}{4\pi \epsilon_0}\frac{e^2}{m_ec^2}
\end{equation}
where $e$ is the electron charge and $\epsilon_0$ is the permittivity of free space. Higher order corrections to the formula exist, but were not used to calculate  $\frac{dE}{dx}$. 

The MIP in a detector material is determined by the energy loss, $\frac{dE}{dx}$, and the thickness of the active material in the detector:
\begin{equation}
MIP_{material}(x) = \frac{dE}{dx} \cdot x
\label{eqn:MIP}
\end{equation}
where $x$ is the thickness of the active area.

The number of ionization pairs subsequently produced is proportional to the energy of the incident particle, $E_0$, divided by the mean energy to produce an ionization pair, $E_i$, as shown in Equation~\ref{eqn:Nions}:
\begin{equation}
N_{ion~pairs} = E_0/E_i
\label{eqn:Nions}
\end{equation}
where $E_i$ is roughly proportional to the band gap energy, $E_g$: 
\begin{equation}
E_i \propto E_g~.
\label{eqn:Eiprop}
\end{equation}
Table~\ref{tab:bandgap} shows some known values of $E_g$ and $E_i$. A linear fit was performed to the data in Table~\ref{tab:bandgap} to arrive at an approximation for $E_i$:
\begin{equation}
E_i \approx 2.0877 \cdot E_g + 1.2122~.
\label{eqn:Ei}
\end{equation}
Then, the number of electron-hole pairs generated by a MIP is given by:
\begin{equation}
N_{ion~pairs} = \frac{\frac{dE}{dx} \cdot x}{E_i}~.
\label{eqn:ions_from_MIP}
\end{equation}
\begin{table}[ht]
  \centering
  \begin{tabular}{c|cc}
  \hline
  Material	& $E_g$ (eV)  	& $E_i$ (eV)	\\
  Ge		& 0.67		& 2.96		\\
   Si		& 1.11		& 3.62		\\
  CdTe	& 1.4			& 4.43		\\
  GaAs	& 1.43		& 4.2			\\
  diamond & 5.5		& 13			\\
    \hline
 \end{tabular}
   \caption{The bandgap energy, $E_g$, and mean energy to produce an electron-hole pair, $E_i$, for several materials. }
   \label{tab:bandgap}
\end{table}

Table~\ref{tab:materials_charge} shows the charge collection properties of several types of semiconducting materials. The more traditional silicon, germanium, and diamond are compared with potential TF material candidates. The list is not exhaustive of TF materials~\cite{muon_stop, sylvaco, palmer, twiki}, but represents a starting point for consideration.
\begin{table}[ht]
  \centering
  \begin{tabular}{c|cccccc}
  \hline
  	~ 	& 	$Z$	 	&	$\rho$ 	&$\frac{-dE}{dx}$ &	MIP	&  $E_i$ 	&  $<N_{e-h~pairs}>$	\\	
  Material & ~ 			&  (g/cm$^3$) 	&  [MeV/(g/cm$^2$)] &  in 10$\mu$m (keV) &  (eV) &  in 10 $\mu$m \\
  \hline
B		&  	5 		&	2.370	& 1.623	&	3.85		&		&		\\
Diamond	& 	6		&	3.51		& 1.78	&	6.25		&	13	&	0.5k	\\
Si		& 	14		&	2.329	& 1.664	&	3.9	 	&	3.62 & 	1.1k     \\
S		& 	16		&	2.00		& 1.652	&	3.30		&	6.64*&	0.5k	\\
Zn		& 	30		&	7.133	& 1.411	&	10.06	&	8.1*	&	1.2k	\\
Ga		& 	31		&	5.904	& 1.379	&	8.14		&		&		\\	
Ge		& 	32		&	5.323	& 1.370	&	7.29		&	2.96	&	2.5k	\\
As		& 	33		&	5.730	& 1.370	&	7.85		&		&		\\
Cd		& 	48		&	8.650	& 1.277	&	11.05	&		&		\\
I		& 	53		&	4.930	& 1.263	&	6.23		&		&		\\	
Pb		& 	82		&	11.350	& 1.122	&	12.73	&		&		\\	
~		& 	~		&	~		& ~		&	~		& ~		&	~	\\				
CdTe	& 	50		&	6.2		&  1.26	&	7.81		&	4.43	&	1.8k	\\	
CdS		& 	32		&	4.8		&  4.0*	&	19.08	& 6.49*	&	2.9k	\\	
PbS		& 	49		&	7.6		& 6.2*	&	46.8		& 1.98*	&	23.6k\\	
ZnO		& 	19		&	5.6		& 4.4*	&	24.8		& 8.25*	&	3.0k	\\
GaAs	& 	32		&	5.32		&  1.4	&	7.45		&	4.2	&	1.8k	\\
InP		& 	32		&	4.97		& 4.0*	&	20.5		&	4.2	&	4.8k	\\
HgI		& 	66.5		&	6.4		& 5.6*	&	35.8		&	4.3	&	8.3k	\\
InSb		& 	50		&	5.78		& 4.9*	&	28.1		& 1.57*	&	17.9k\\
InAs		& 	41		&	5.67		& 4.7*	&	26.8		& 1.94*	&	13.8k\\
HgTe	& 	66		&	8.1		& 6.7*	&	54.7		&		&		\\
CdZnTe	& 	43.3		&	6		& 5.0*	&	29.8		&	4.7	&	6.3k	\\
IGZO		& 	29.5		&	6		&		&			& 7.58*	&		\\

  \hline
 \end{tabular}
   \caption{Charge collection properties of various semiconducting materials. ~\cite{muon_stop, sylvaco, palmer, twiki} *dE/dx and $E_i$ values were calculated using Equation~\protect\ref{eqn:dEdx} and \protect\ref{eqn:Ei} respectively. }
   \label{tab:materials_charge}
\end{table}
Lead Sulfide, PbS, and Indium Antimonide, InSb, materials show promise for use in TF detectors. It is estimated they will generate over 17,000 and 23,000 electrons in 10 $\mu$m for the signal. Table~\ref{tab:materials_time} lists the material properties relating to the charge collection time. 


\subsubsection{Signal Timing}\label{sec:timing}

The signal collection time is important depending on the type of HEP experiment. The LHC is capable of bunch-crossings every 25 ns or a frequency of 40 MHz. Ideally, a signal could be read out in that amount of time or less to be ready for the next possible interaction. For example, silicon has a characteristic collection time of 34 ns for holes and 11 ns for electrons in 300 $\mu$m thick detector at a bias voltage of 60 V. However, there are also applications for slower read-out as well. Therefor, the selected target for the signal collection time is in the range 10-100 ns. 

The collection time depends strongly on the electrical properties of the material and the distance the charge carriers travel. The velocity, $v$, for a charge carrier is proportional to the mobility, $\mu$, times the electric field, $E$:
\begin{equation}
v = \mu E.
\end{equation}
The electric field will depend on the detector thickness, applied bias voltage, and the effective doping concentration, which may be manipulated in the detector design. The mobility covers a wide range for different materials as shown in Table~\ref{tab:materials_time}. It is possible to select materials with mobilities similar to silicon and even larger for faster signals.

\begin{table}[ht]
  \centering
  \begin{tabular}{c|cc}
  \hline
  Material & $\mu_e $ $(\frac{cm^2}{V \cdot s})$ & $\mu_h $ $(\frac{cm^2}{V \cdot s})$ \\
  \hline
Diamond	& 1800	& 1200	\\
Si		& 1350	& 480	\\
	
CdTe	& 1050	& 100	\\
CdS		&  340	&  50		\\
PbS		&  600	&  700	\\
ZnO		&  130	& 		\\
IGZO		&  15		&  0.1	\\
GaAs	&  8000	&  400	\\
InP		&  4600	&  150	\\
HgI		&  100	&  4		\\
InSb		&  78000	&  750	\\
InAs		&  33000 	&  460	\\
HgTe	&  22000	&  100	\\
CdZnTe	&  1350	&  120	\\
 \hline
 \end{tabular}
   \caption{Electron and hole mobilities of various semiconducting materials. ~\cite{muon_stop, sylvaco, palmer, twiki}}
   \label{tab:materials_time}
\end{table}

\subsection{Layout}\label{sec:layout}
A TF sensor will consist of a flexible substrate material approximately 1 $\mu$m thick, layered with the bulk detector material on the order of 10-50 $\mu$m. It is foreseen that an additional source of amplification will be needed for the detection material. An avalanche diode could be implanted into each pixel and serve to amplify the signal. An integrated charge amplifier and shaper with discrimination logic would be placed in an insulating well separated from its neighbors. The signals for each pixel will then be traced out in additional insulated layers. Pixel sizes can easily be scaled under 10 $\mu$m if necessary.

\begin{figure}[htb]
\centering
\includegraphics[width=0.3\textwidth]{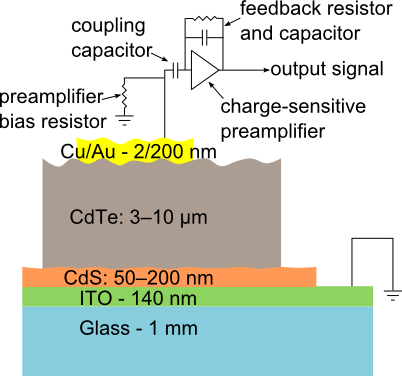}
\includegraphics[width=0.6\textwidth]{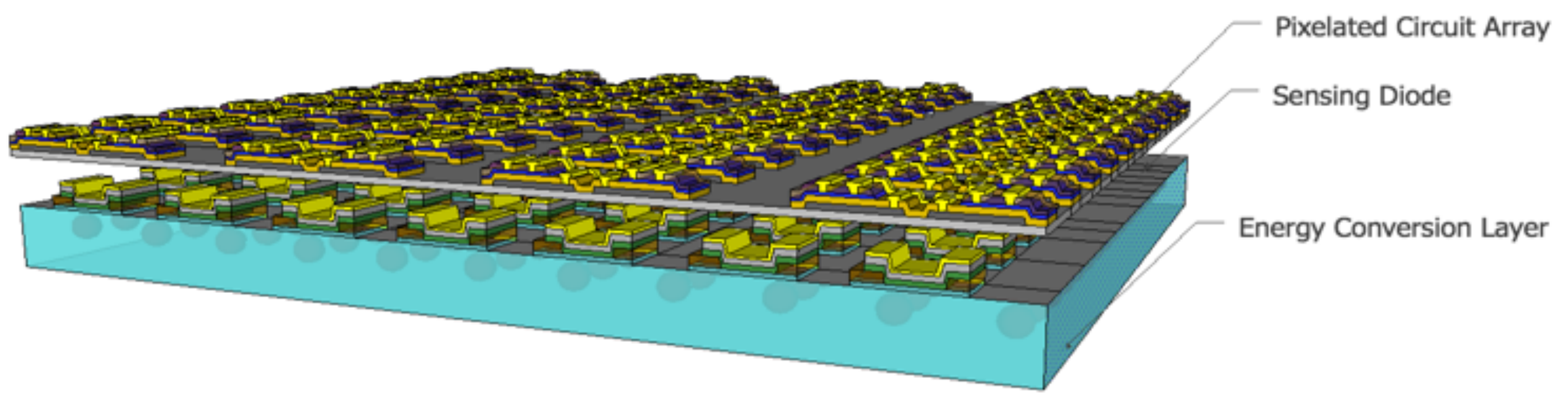}
\caption{The cross-section of a detection diode is shown on the left with typical materials and thicknesses. An example of an array of a pixelated detector with integrated circuits that can be produced with Thin Films.}
\label{fig:layout}
\end{figure}

The total thickness of the detector is estimated to be approximately 50 $\mu$m. The average density is estimated at roughly 6 g/cm$^3$ and would be approximately 0.01 radiations lengths per layer. Figure~\ref{fig:layout} shows a diagram of a possible layout.

\subsection{Geometries}
\begin{figure}[htb]
\centering
\includegraphics[width=0.3\textwidth]{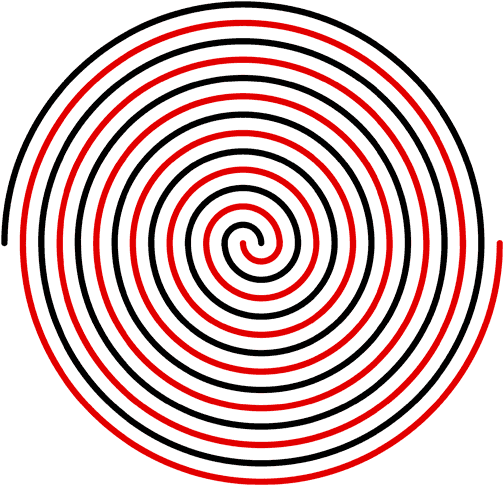}
\caption{An example of an innovative geometric layout for two detectors that could be possible with a flexible substrate. }
\label{fig:design}
\end{figure}

TF technology can generally be constructed on single sheets as large as 1-2 m$^2$. The radius of curvature is much less than a typical beam pipe. A single sheet can form a cylinder around the beam pipe leaving only one seam of dead material. The flexible nature of the sensor allows for many more types of geometries that could be adapted to the needs of a given experiment. Figure~\ref{fig:design} illustrates a few possibilities for detector geometries. 
  
\section{Radiation Properties of TF}

In particle physics experiments, the detector elements and its associated
electronics 
will be subjected to background radiation. The largest radiation levels
are found near the accelerator collision points. 
All charged particles produced there, 
mesons in the vast majority, will deposit ionization dose and induce
displacement damage. The development of devices and electronics that
can withstand large doses and are relatively immune to displacement damage
is one the biggest challenges in developing  tracking detectors. 
Typical 1 MeV neutron equivalent fluences are on the order of 10$^{13}$-10$^{17}$ 1 MeV $n_{eq}$/cm$^2$. 

Thin film electronics when compared to traditional silicon devices do
not absorb large amounts of energy. Typical device thicknesses are less than
20~nm which makes the deposition of energy small. These thicknesses also
disfavor the trapping of charge in the insulating layers. The net result is
that they are inherently radiation resistant. As the interest in the use
of TF based electronics grow in various fields of application, 
data on the effects of radiation in their performance has become available.
Alvarado, et al., exposed TF transistors to a $^{60}Co$ to doses up to
10~Mrad observing no changes in the transistor parameters. \cite{Zhao, Ramirez} This result is
encouraging as typical test doses are well above the values found at
accelerators. 

Displacement damage caused by hadrons in the crystalline structure is less
known for TFs. This is one of the areas that will require detailed study. 
However, we expect that this type of effect to be small or negligible. 
There is no observable effects in CMOS devices even for the small
feature size devices. However, the difference here is the choice of
materials that can lead to new effects and need to be determined
experimentally. 

Finally, single event upsets needs to be considered in case logic devices
such as switches are to be implemented near the detector. TF devices
operate at very low voltages requiring less charge to be switched. On the
other hand the very nature of the geometry favors less charge creation
and therefore a detailed simulation and experimental study should
be carried out. One difference between TF and silicon based devices is
that the observation of volumetric effects such as funneling is not expected in TFs. Thus
comparatively, we should expect less sensitivity to SEU mostly because
charge, larger than $Q_{crit}$, must be created near the device
sensitive areas to induce an SEU.

\section{TF Detectors}
\subsection{TF Neutron Detector}
\begin{figure}[htb]
\centering
\includegraphics[width=0.8\textwidth]{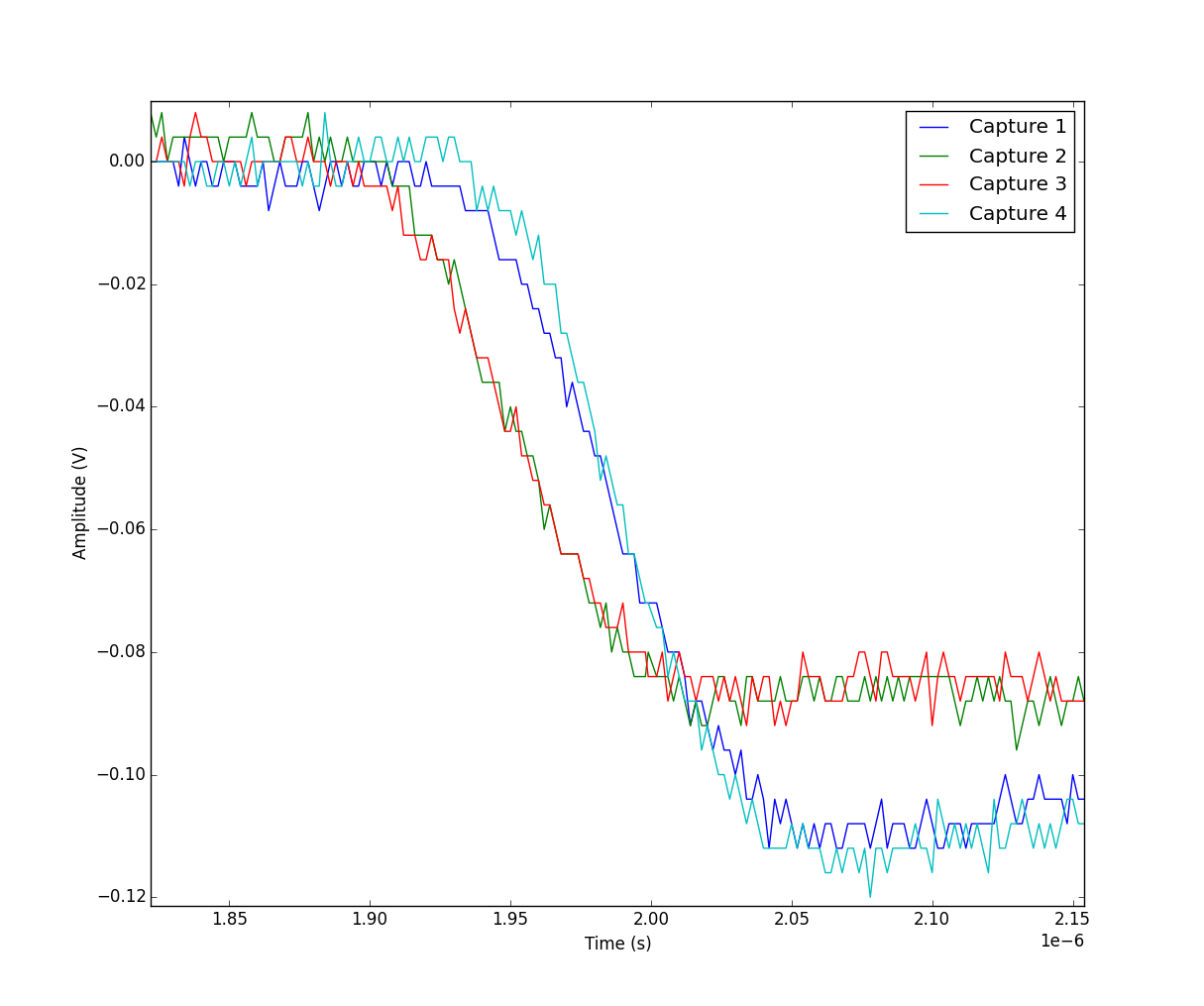}
\caption{The inverted signal of the rising edge from an alpha particle from a $^{210}$Po source in a CdTe detector 6 $\mu$m active thickness with 1 V bias is shown. The rising edge is on the order of 100 ns.  }
\label{fig:sample_signal_peak}
\end{figure}

A successful TF neutron detector was built and tested using CdTe detection material and a Boron conversion layer \cite{UTCdTeSensor, UTdiodeThickness} demonstrating the feasibility of this technology as a particle detector. The detection layer was 6 $\mu$m thick with 4 $\mu$m thick active region. It was exposed to a $^{210}$Po alpha radiation source with particle energy approximately 3500 keV. The collection efficiency was $>$~80\% with a bias voltage of 2 V. The reduction in efficiency was greatly due to the electrode spot size and the probes between the source and electrodes. The full width half maximum (FWHM) energy resolution was approximately 27\% to 41\%. Figure~\ref{fig:sample_signal_peak} shows a typical signal from the $^{210}$Po source using off-detector preamp and shaping amplifier. The collection time is on the order of 100 ns.

The amount of charge collected and the charge collection time agree with those predicted for CdTe in Tables~\ref{tab:materials_charge} and~\ref{tab:materials_time}. 

\section{Summary}
The potential of TF technology and materials to design a charged particle detector for use as a tracking detector in HEP experiments was explored. PbS and InSb materials show promising levels of collected charge and signal time. These materials may meet the charge collection amplitude and time targets of 1,000-10,000 electrons and 10-100 ns, respectively. A successful neutron detector with CdTe was previously measured using a $^{210}$Po source. The results were found to be consistent with expectation. There are many potential benefits for using TF technology including flexible detectors, low cost, large area, low material budget, low power, less cooling, and high pixel granularity.


\acknowledgments

Acknowledgments.


\begin{thebibliography}{9}



\bibitem{street}
Robert A. Street,
\emph{Thin Film Transistors},
\emph{Advanced Materials} \textbf{Volume 21, Issue 20,} (2009) p 2007-2022.

\bibitem{gnade}
Bruce Gnade, 
\emph{Thin-Film Semiconductor Technology Applied to Large Area Radiation Detectors}, 
\emph{CIRMS Conference}, \href{http://www.cirms.org/pdf/2012_conference_pdf/Tuesday\%20Morning/cirms2012\%20Gnade.pdf} (2012).

\bibitem{mejia}
I. Mejia, M. Estrada, M. Avila,
\emph{Improved upper contacts PMMA on P3HT PTFTS using photolithographic processes},
\emph{Microelectronics Reliability} \textbf{vol 48,} (2008) p 1795-1799.

\bibitem{thickFilms}
Q. Jiang, et al.,
\emph{Growth of Thick Epitaxial CdTe Films by Close Space Sublimation},
Journal of ELECTRONIC MATERIALS, \textbf{Vol. 38, No. 8}, DOI: 10.1007/s11664-009-0808-1, 2009.

\bibitem{muon_stop}
D. E. Groom, N. V. Mokhov, and S. Striganov,
\emph{Muon Stopping Power and Range},
\emph{Atomic Data and Nuclear Data Tables}, \textbf{Vol. 76, No. 2}, (July 2001).

\bibitem{sylvaco}
ATLAS Users Manual, Sylvaco, 
\href{https://dynamic.silvaco.com/dynamicweb/jsp/downloads/DownloadManualsAction.do?req=silen-manuals&nm=atlas} (2014).

\bibitem{palmer}
D W Palmer, 
\emph{The Semiconductors-Information Web-site }
www.semiconductors.co.uk,
\href{www.semiconductors.co.uk} (March 2008).

\bibitem{twiki}
Wikipedia, The free encyclopedia, http://en.wikipedia.org/wiki/TWiki,
\href{http://en.wikipedia.org/wiki/TWiki} (2014).


\bibitem{Zhao}
Dalong Zhao, D. A. Mourey, T. N. Jackson, 
\emph{Gamma-ray irradiation of ZnO thin film transistors and circuits},
Device Research Conference (DRC), DOI: 10.1109/DRC.2010.5551979, (June 2010) p~241-242. 

\bibitem{Ramirez}
J. I. Ramirez, Y. V. Li, H. Basantani, T. N. Jackson,
\emph{Effects of gamma-ray irradiation and electrical stress on ZnO thin film transistors},
71st Annual Device Research Conference, DOI: 10.1109/DRC.2013.6633848, (June 2013) p 171-172. 


\bibitem{UTCdTeSensor}
J. W. Murphy, L. Smith, J. Calkins, G. R. Kunnen, I. Mejia, K. D. Cantley, R. A. Chapman, J. Sastr$\acute{e}$-Hern$\acute{a}$ndez, R. Mendoza-P$\acute{e}$rez, G. Contreras-Puente, D. R. Allee, M. Quevedo-Lopez, and B. Gnade,
\emph{Thin film cadmium telluride charged particle sensors for large area neutron
2 detectors}, submitted to APPLIED PHYSICS LETTERS \textbf{105},  (2014).

\bibitem{UTdiodeThickness}
John W. Murphy, George R. Kunnen, Israel Mejia, Manuel A. Quevedo-Lopez, David Allee et al.,
\emph{Optimizing diode thickness for thin-film solid state thermal neutron detectors},
Appl. Phys. Lett. \textbf{101}, 143506, doi: 10.1063/1.4757292 (2012).




\end{thebibliography}
\end{document}